\documentclass[iop]{emulateapj} 

\usepackage{pifont} 
\shorttitle{Zodiacal Light Spectrum with CIBER LRS}
\shortauthors{Tsumura et al.}
\newcommand{\myemail}{{\it tsumura@ir.isas.jaxa.jp}}

\begin{document}
\title{Observations of the Near-Infrared Spectrum of the Zodiacal Light with CIBER}
\author{K. Tsumura\altaffilmark{1}, J. Battle\altaffilmark{2}, J. Bock\altaffilmark{2,3}, A. Cooray\altaffilmark{4}, V. Hristov\altaffilmark{3}, B. Keating\altaffilmark{5},
        D. H. Lee\altaffilmark{6},\\ L. R. Levenson\altaffilmark{3}, P. Mason\altaffilmark{3}, T. Matsumoto\altaffilmark{1,7}, S. Matsuura\altaffilmark{1}, 
        U. W. Nam\altaffilmark{6}, T. Renbarger\altaffilmark{5}, I. Sullivan\altaffilmark{3},\\ K. Suzuki\altaffilmark{8}, T. Wada\altaffilmark{1}, 
        and M. Zemcov\altaffilmark{2,3}}
\affil{\myemail}
\slugcomment{Accepted by the Astrophysical Journal}
        
\altaffiltext{1}{Department of Infrared Astrophysics, Institute of Space and Astronautical Science (ISAS), Japan Aerospace Exploration Agency (JAXA), Sagamihara, Kanagawa 252-5210, Japan}
\altaffiltext{2}{Jet Propulsion Laboratory (JPL), National Aeronautics and Space Administration (NASA), Pasadena, CA 91109, USA}
\altaffiltext{3}{Department of Astronomy, California Institute of Technology, Pasadena, CA 91125, USA}
\altaffiltext{4}{Center for Cosmology, University of California, Irvine, Irvine, CA 92697, USA}
\altaffiltext{5}{Department of Physics, University of California, San Diego, San Diego, CA 92093, USA}
\altaffiltext{6}{Korea Astronomy and Space Science Institute (KASI), Daejeon 305-348, Republic of Korea}
\altaffiltext{7}{Department of Physics and Astronomy, Seoul National University, Seoul 151-742, Republic of Korea}
\altaffiltext{8}{Instrument Development Group of Technical Center, Nagoya University, Nagoya, Aichi 464-8602, Japan}

\begin{abstract}
Interplanetary dust (IPD) scatters solar radiation which results in the zodiacal light that dominates the celestial diffuse brightness at optical and near-infrared wavelengths.  
Both asteroid collisions and cometary ejections produce the IPD, but the relative contribution from these two sources is still unknown.
The Low Resolution Spectrometer (LRS) onboard the Cosmic Infrared Background ExpeRiment (CIBER) observed
the astrophysical sky spectrum between 0.75 and 2.1$\, \mu$m over a wide range of ecliptic latitude. 
The resulting zodiacal light spectrum is redder than the solar spectrum, and shows a broad absorption feature, previously unreported, at approximately 0.9$\, \mu$m, 
suggesting the existence of silicates in the IPD material. 
The spectral shape of the zodiacal light is isotropic at all ecliptic latitudes within the measurement error.
The zodiacal light spectrum, including the extended wavelength range to 2.5$\, \mu$m using Infrared Telescope in Space (IRTS) data, 
is qualitatively similar to the reflectance of S-type asteroids.
This result can be explained by the proximity of S-type asteroidal dust to Earth's orbit, and the relatively high albedo of asteroidal dust compared with cometary dust.
\end{abstract}


\section{INTRODUCTION}
The astrophysical sky brightness is dominated by the zodiacal light which comprises scattered sunlight by interplanetary dust (IPD) in the optical and near-infrared, 
and thermal emission from the same IPD in the mid- and far-infrared.
Historically, extensive zodiacal light observations in the scattered sunlight regime were conducted from ground-based observations at high altitude sites in the 1960's and 1970's \citep{Levasseur80}.
However, these were restricted to optical wavelengths because atmospheric OH airglow is much brighter than the zodiacal light in the near-infrared. 
Space-based platforms get above this contamination and provide precision measurements of the diffuse sky brightness at near-infrared wavelengths
\citep{Murdock85, Berriman94, Matsuura95, Matsumoto96}.
Based on measurements from DIRBE on {\it COBE}, phenomenological models for the IPD distribution have been constructed \citep{Kelsall98, Wright98} 
to estimate the zodiacal foreground for measurements of the extragalactic background light (EBL; \citet{Wright98, Hauser98, Matsumoto05}).

A continuous supply of IPD particles is necessary to replenish the zodical dust cloud,
as they either fall into the Sun through Poynting-Robertson drag or leave the inner solar system due to radiation pressure.
However, it still remains unclear whether the dominant source of IPD is asteroids or comets.
Recent dynamical analyses suggest that IPD arises from a cometary origin \citep{Liou95, Ipatov08, Nesvorny09};
dust trails of short-period comets show that the mass loss from cometary nuclei amounts to a third to a half of the required supply of dust \citep{Ishiguro09}.
Additionally, particles captured in the stratosphere suggest that IPD from asteroids and comets occur with comparable abundance 
with cometary particles being slightly more common \citep{Schramm89}.
On the other hand, three dust bands in the IPD cloud were found by {\it IRAS} \citep{Low84} and their parent bodies were identified as being associated with certain asteroid families \citep{Dermott84, Nesvorny03}.
Age estimates of the last collisions in these families are consistent with such an association \citep{Nesvorny03}.
The discovery of these dust bands is strong evidence that there is an important asteroidal contribution to IPD.

\begin{table*}
\caption{Summary of the CIBER Observation Targets\label{flight}}
\begin{center}
\vskip -0.5cm
\begin{tabular}{ccccccccccc}
\tableline\tableline\noalign{\smallskip}
Target & Time [sec] & Altitude [km] & R.A. & Decl. & $\lambda$ & $\beta $ & $l$ & $b$ & Zenith Angle & Solar Elongation \\
\noalign{\smallskip}\tableline\noalign{\smallskip}
Elat-10 & 90 - 109 & 139 - 172 & 234.05 & -8.32 & 233.76 & 10.71 & 357.23 & 36.63 & 44.4 & 102.0 \\
Elat-30 & 127 - 142 & 199 - 220 & 222.75 & 20.56 & 212.82 & 35.10 & 25.90 & 61.96 & 74.7 & 117.1 \\
Bo\"otes-A & 162 - 250 & 245 - 310 & 218.63 & 34.84 & 201.23 & 46.28 & 58.59 & 66.68 & 84.5 & 119.3 \\
Bo\"otes-B & 262 - 358 & 314 - 295 & 217.33 & 33.39 & 202.23 & 47.28 & 55.43 & 68.02 & 84.5 & 119.3 \\
NEP & 370 - 426 & 288 - 232 & 270.29 & 65.88 & - & 90 & 95.60 & 29.69 & 42.7 & 89.9 \\
SWIRE ELIAS-N1 & 440 - 490 & 215 - 135 & 242.56 & 55.21 & 208.40 & 72.63 & 85.21 & 44.66 & 59.3 & 100.6 \\
\noalign{\smallskip}\tableline\noalign{\smallskip}
\end{tabular}
\tablecomments{The time column denotes the elapsed time after launch and ($\lambda $,$\beta $) and ($l$,$b$) denotes ecliptic and Galactic coordinates, respectively.
All angles are given in degrees.}
\end{center}
\end{table*}

Because materials likely to be present in the IPD have distinct spectral features in the near-infrared,
spectroscopic measurement of the zodiacal light may help to determine the composition of IPD particles.
Measurements based on bandpass photometry indicate that the color of the zodiacal light is redder than the solar spectrum \citep{Matsuura95, Matsumoto96}, 
which implies that dust particles larger than 1$\, \mu$m are mainly responsible for the zodiacal light \citep{Matsuura95}.
\citet{Matsumoto96} found a silicate-like feature at around 1.6$\, \mu$m from Infrared Telescope in Space (IRTS) data, 
and suggested that dust from S-type asteroids are responsible for the zodiacal light.
In addition, a silicate feature was found in mid-infrared spectrum of the zodiacal light by IRTS \citep{Ootsubo98}, 
{\it Infrared Space Observatory} ({\it ISO}; \citet{Reach03}), and {\it AKARI} \citep{Ootsubo09}.

\section{THE COSMIC INFRARED BACKGROUND EXPERIMENT}
\subsection{Overview of CIBER}
The Cosmic Infrared Background ExpeRiment (CIBER) is a rocket-borne instrument to search for signatures from first-light galaxies 
in the near-infrared extragalactic background (\citealt{Bock06}, \citealt{Zemcov10}). 
CIBER consists of four optical instruments: two wide-field imagers \citep{Sullivan10}; 
a Narrow Band Spectrometer (NBS; \citealt{Renberger10}); and a Low Resolution Spectrometer (LRS; \citealt{Tsumura10}).
The LRS data from CIBER's first flight are presented in this paper.  
These instruments are mounted to a common optical bench which is cooled by liquid nitrogen to reduce thermal instrumental emission.
A Terrier-Black Brant IX rocket carrying the CIBER payload was successfully launched from White Sands Missile Range on 2009 February 25,
achieving an apogee of $\sim$330 km, and providing more than 420 s of astronomical data.  
After the flight, the instrument was successfully recovered for future flights.

\begin{figure}
\begin{center}
\includegraphics[scale=0.3,angle=90]{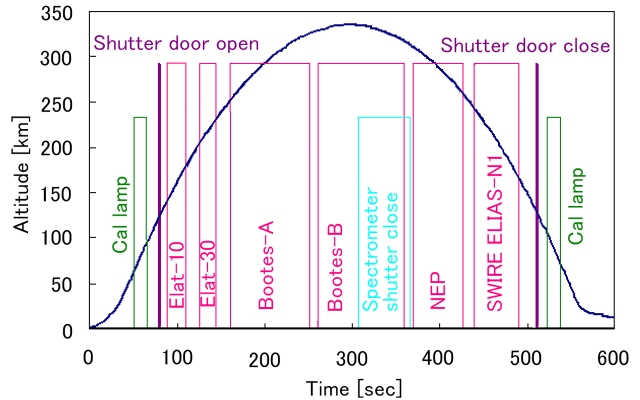}
\end{center}
\vskip -0.2cm
\caption{Instrument altitude and science observation field sequence as a function of time for CIBER's first flight. 
Reference calibration lamps and cold shutters in front of the focal planes were operated in flight to allow calibration transfer and absolute dark level reference.\label{altitude}}
\end{figure}

The choice of our science fields is shown in Figure~\ref{altitude} and Table~\ref{flight}.  
Since the Bo\"otes and SWIRE ELIAS-N1 fields were well studied with {\it Spitzer} \citep{Eisenhardt04, Lonsdale03}
and the north ecliptic pole (NEP) field is well studied with {\it AKARI} \citep{Matsuhara06}, these fields were chosen for the CIBER imaging instruments.
The Elat-10 and Elat-30 fields were chosen specifically to measure the dependence of the zodiacal light intensity on ecliptic latitude, 
while the remaining science fields were chosen at higher ecliptic latitudes to reduce the brightness of the zodiacal foreground for the EBL science.  
However, as the zodical light intensity is never small, the EBL fields can also be used to constrain the zodical light. 
Each of these fields was observed in a series of 25 seconds reset intervals.  
As the Bo\"otes field ancillary coverage is very wide, this field was observed in two pointings, A and B, with each observed for about 90 seconds.  
During the last 50 seconds on Bo\"otes-B, the cold shutters on the both spectrometers were closed to obtain dark frames for absolute spectroscopy. 
The pointing stability in 25 seconds was less than 3 arcsec.

\subsection{The Low Resolution Spectrometer}
The LRS is designed to obtain absolute spectra of the astrophysical sky brightness in the near-infrared.
Although primarily designed to measure the absolute sky brightness from extragalactic sources, 
including first-light sources present during re-ionization, the LRS also measures the zodiacal light foreground. 

The specifications of the LRS are summarized in Table~\ref{LRSspec}.  
The LRS is designed to cover the wavelength range 0.75-2.1$\, \mu$m with a spectral resolution $\lambda /\Delta \lambda$ varying between 15 and 30. 
The LRS employs a 5 cm refractive telescope and has a cold shutter just before the focal plane detector to monitor the detector dark current for absolute spectroscopy.  
The shutter is closed prior to launch, on ascent, during flight, and on descent.  
The LRS optics form an initial focus at a field stop, where a mask with five equally spaced slits is placed. 
The optics then relay the field stop back to a parallel beam where a prism disperses incident light perpendicular to the slit mask direction.  
Finally, this parallel beam is refocused on a 256$\times $256 substrate-removed HgCdTe PICNIC detector array; 
the slit mask is imaged to five 2.8 arcmin $\times $ 5.5 degree strips on the sky at the focal plane. 
A calibration lamp is operated both before and after astronomical observations to confirm the stability of the instrument responsivity during the flight: 
the detector was stable to better than 4\% during the flight.
This calibration lamp also serves as a transfer standard between the flight data and laboratory-based calibration measurements. 

The wavelength and responsivity calibration of the LRS were measured in the laboratory before flight.
For the wavelength calibration, radiation from a halogen lamp was dispersed by a monochromator,
the output of which is fiber coupled to an integrating sphere producing an approximately solar spectrum, aperture-filling input source.
The LRS observes the integrating sphere while the wavelength is stepped from 0.75 to 2.1$\, \mu$m in 0.005$\, \mu$m increments.  
Additionally, the responsivity of the LRS was calibrated using the halogen lamp and an optical/near-infrared tunable laser covering the range from 0.7 to 1.6$\, \mu$m.
Each of these sources is coupled to an integrating sphere and is measured simultaneously by both the LRS and absolutely calibrated detectors (SIRCUS; \citealt{Brown04}).
These measurements are used to accurately calibrate both the absolute responsivity and the channel-to-channel relative responsivity of the LRS.

Further details of the instrument including the calibration and flight performance are summarized in \citet{Tsumura10}.

\begin{table}
\caption{Specifications of the Low Resolution Spectrometer\label{LRSspec}}
\begin{center}
\vskip -0.5cm
\begin{tabular}{ll}
\tableline\tableline\noalign{\smallskip}
Characteristic & Value \\
\noalign{\smallskip}\tableline\noalign{\smallskip}
Optics & 14 lenses, 2 filters, 1 prism, 5 slits \\
Aperture & 50 mm \\ 
F number & 2\\
FOV & 5.5 degrees along a slit \\ 
Pixel size & 1.4 arcmin$\times $1.4 arcmin \\ 
Slit & 2 pixels$\times $236 pixels\\ 
Wavelength range & 0.75-2.1$\, \mu$m \\
Wavelength resolution & $\lambda /\Delta \lambda $=15-30 \\ 
Optics efficiency & 0.8 \\
Detector & 256$\times $256 substrate removed PICNIC \\ 
Detector QE & 0.9 \\ 
Dark current & ƒ0.6 $e^{-}$/s \\
Readout noise & ƒ26 $e^{-}$ CDS \\
\noalign{\smallskip}\tableline\noalign{\smallskip}
\end{tabular}
\end{center}
\end{table}

\section{DATA REDUCTION}
\subsection{Obtained Images}
\begin{figure}[hbtp]
\begin{center}
\includegraphics[scale=0.3,angle=90]{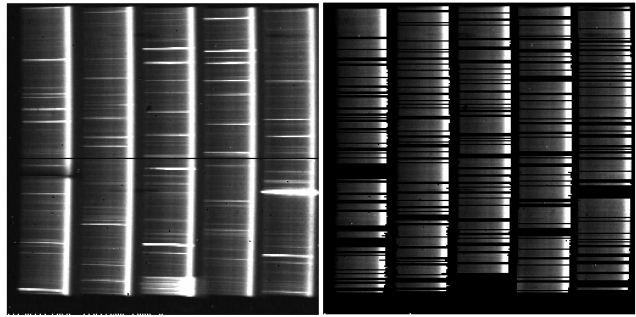}
\end{center}
\vskip -0.3cm
\caption{Image of the NEP field obtained during 25 seconds of integration of the LRS during CIBER's first flight.
The dispersion direction is horizontal and the wavelength increases from left to right in the images, 
while the large intensity in the five perpendicular strips is caused by scattered thermal emission from the hot rocket skin.
The central slit has a small notch located at the bottom center to facilitate laboratory testing; 
the edge of the array is not illuminated by the optics and can be used as a monitor of diffuse stray light falling on the detector. 
The left hand panel includes the raw spectra of stars which happened to fall on the slits as horizontal bands of emission, 
and the detected stars are masked as shown at the right hand panel.
The averaged spectrum of the diffuse sky is obtained by combining these source-masked spectra.\label{NEP}}
\vskip 0.5cm
\end{figure}

Figure~\ref{NEP} shows an LRS image obtained during CIBER's first flight. 
The five vertical slits are dispersed horizontally so that the wavelength increases from left to right in this image.
The horizontal lines are caused by stars which happened to illuminate the slit,
while the five strong vertical lines are caused by stray thermal emission from the hot rocket skin prominent at long wavelengths (see Section \ref{thermalstray_section}).
The central slit has a small notch located at the bottom center used to facilitate laboratory testing. 
A masked region exists at one edge of the array which can be used to monitor the presence of diffuse stray light.

\subsection{Thermal Stray Light}\label{thermalstray_section}
The five strips visible at longer wavelengths in Figure~\ref{NEP} are caused by bright thermal emission from the sounding rocket skin 
which was heated by air friction during the powered ascent.  
This thermal emission couples to the LRS optics via diffuse scattering from the instrument baffle placed before the first aperture.
By fitting a blackbody curve to the thermal spectrum in the LRS images taken during the time when the shutter door was closed,
the temperature of the rocket skin is estimated to have peaked at $\sim$400 K in flight. 
Spillover from this stray thermal emission from the adjacent slit contaminated the short wavelengths regions ($<$0.82$\, \mu$m) of the four rightmost slits, 
so only data from the leftmost slit are used to derive the four spectral data points shorter than 0.82$\, \mu$m.
Contamination by longer wavelength spillover within each slit region was measured after the flight to be less than 0.4\% for $\lambda <$1.4$\, \mu$m \citep{Tsumura10}, 
which is negligibly small for the science described in this paper.  
In addition, we investigated the masked region of the astronomical data and found a low photo current level ($<$2 e$^-$/sec corresponding to $<$100 nW/m$^2$/sr),
a small residual which we subtracted.
There is no evidence that stray thermal emission affects the results presented here at shorter wavelengths, and we simply omitted data at $\lambda >$1.8$\, \mu$m.

\subsection{Airglow Emission}\label{airglow_section}
\begin{figure}
\epsscale{0.9}
\plotone{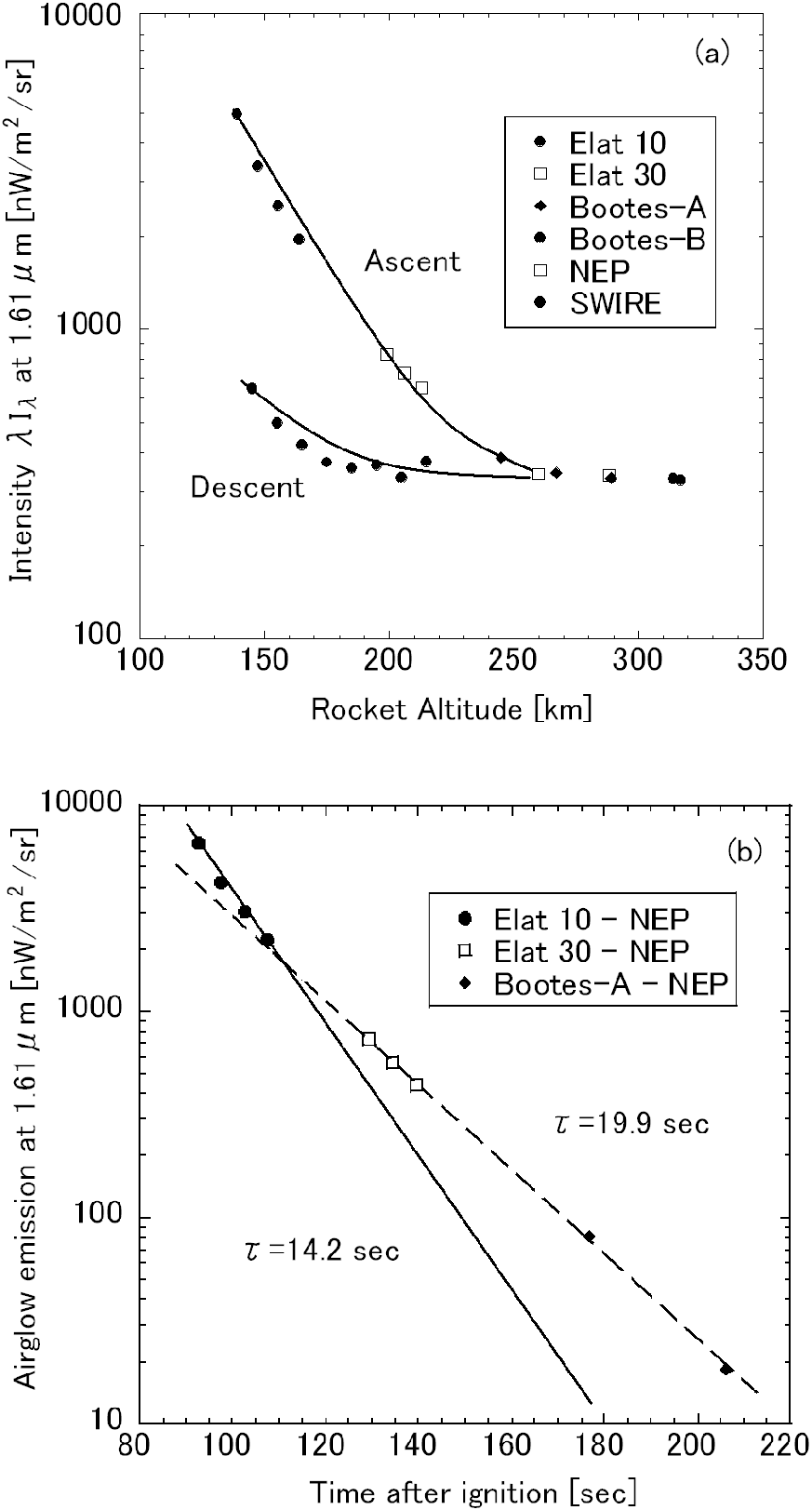}
\caption{(a) Brightness of the airglow emission as a function of instrument altitude at 1.61$\, \mu$m where the airglow emission is brightest.  
The Elat-10, Elat-30, and SWIRE data are separated into several subsets to allow differentiation of the time and altitude dependence of the airglow signal. 
The airglow emission appears to be negligible at altitudes greater than 250km.
(b) Brightness of the airglow emission at 1.61$\, \mu$m which scales exponentially with time during the ascent phase of the flight.  
A similar exponential variation is observed on descent but with a smaller amplitude.  
The decay timescale $\tau$ was 14.2 seconds from only the Elat-10 data, while we obtained $\tau$=19.9 seconds for the entire data sequence.  
We used $\tau$=14.2 seconds for deriving the absolute brightness of the airglow emission at Elat-10 field. \label{airglow}}
\end{figure}

A second emissive component not associated with the astronomical sky is observed to have exponential time and altitude dependence as shown in Figure~\ref{airglow}. 
Based on its spectrum, which has prominent emission features at 1.1 and 1.6$\, \mu$m, we attribute this component mainly to OH molecules 
from either exospheric atmospheric airglow or dissociated water vapor which outgasses from the payload early in the flight.  
The spectral shape of the airglow emission is observed to be the same during ascent and descent, 
but as shown in Figure~\ref{airglow} the absolute brightness of the airglow emission on ascent is much brighter; 
this suggests an enhancement by dissociated water outgassing from the payload early in the flight.
This airglow emission is bright in the Elat-10 and Elat-30 field data observed early in the flight, and is essentially negligible for the remaining fields; 
Figure~\ref{airglow} shows the brightness of the airglow emission at 1.61$\, \mu$m in each field derived by subtracting the NEP brightness.

As the Elat-10 field was observed earliest in the flight and at the lowest altitude where we would expect this airglow emission to be brightest, 
these data are used to derive a template spectrum of the contaminant. 
In order to subtract the effect of the airglow emission, a spectral model was derived.
The total signal $I$ can be modeled as 
\begin{equation}
\label{eq:airglow}
I = A(\lambda )+B(\lambda )\exp(-t/\tau),
\end{equation}
where $A(\lambda )$ is the astrophysical sky brightness, $B(\lambda )$ is the airglow emission, and $\tau$ is the timescale of the decay.
Here we employ a fit to airglow in time not in rocket's altitude because the airglow emission on ascent is dominated by the outgassing,
and the outgassing should follow the exponential decay in time rather than in altitude.  
To increase the constraining power of the data, the Elat-10 data is reduced into four and Elat-30 into three sequential spectra.  
Because the sky the LRS observes during the observation of a single field is static, any change in the mean absolute brightness must be caused by time variation in the airglow component.
The spectral shape of the airglow component $B(\lambda )$ is derived by a difference of the brightness 
between the first block and the last block in Elat-10.
The decay timescale $\tau$ was also derived with a fit to the data in Figure~\ref{airglow} and we obtained $\tau=14.2$ seconds from Elat-10 data.
Combining the differencing and the obtained decay timescale, we derived the absolute brightness of the airglow emission at Elat-10.

\begin{figure}
\epsscale{0.98}
\plotone{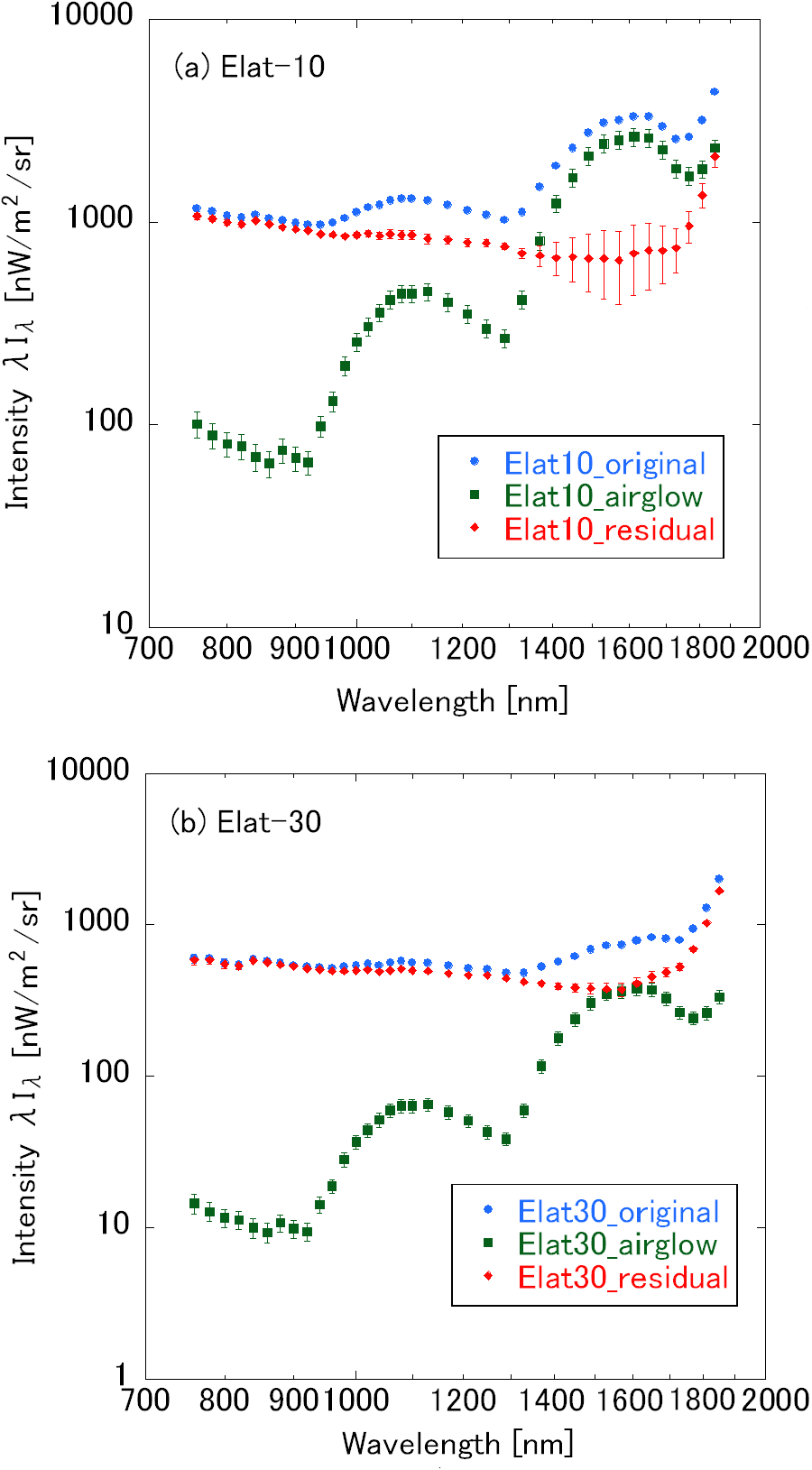}
\caption{Subtraction of the airglow emission for the Elat-10 and Elat-30 fields observed early in the ascent phase where airglow is brightest.
The blue points are \textquotedblleft original\textquotedblright\ spectra of the diffuse sky from about 20 seconds integration, 
which include both the airglow emission and astrophysical sky brightness. 
The green points are spectra of the airglow emission at each field derived by the method described in Section \ref{airglow_section}.
The red points show the residual spectra obtained after subtracting the airglow emission.\label{subtract}}
\end{figure}

The absolute brightness of the airglow emission at Elat-10 derived using this method was checked by comparing the color of the residual spectrum 
between 1.61$\, \mu$m (maximum airglow emission) and 1.29$\, \mu$m (local minimum airglow emission) to the spectrum of the other fields.  
We confirmed that the color of the residual spectrum of Elat-10 was consistent with the color of the NEP and Bo\"otes spectra to within 5\%.  
The color of the SWIRE ELIAS N-1 field was 10\% higher than the others 
because SWIRE was observed at the end of the flight sequence and the airglow emission began to rise again on descent. 
Therefore, we attribute a 5\% error to our estimation of the airglow emission based on the color of the residuals.  
The absolute brightness of the airglow emission at the Elat-30 field was estimated by the same color estimation method and subtracted,
that is, the airglow intensity at Elat-30 was set to make the color of the residual spectrum between 1.61$\, \mu$m and 1.29$\, \mu$m same as NEP and Bo\"otes.  
Figure~\ref{subtract} shows the pre- and post-subtraction spectra with the airglow emission.
We are confident that our main scientific result on the detection of the 0.9$\, \mu$m absorption feature is not strongly affected by airglow emission. 

\begin{table*}
\caption{Comparison of Total and Integrated Galactic Star Light at {\it J}-band\label{LRSstar}}
\begin{center}
\vskip -0.5cm
\begin{tabular}{cccccc}
\tableline\tableline\noalign{\smallskip}
Sky Brightness & Elat-10 & Elat-30 & Bo\"otes & NEP & SWIRE \\
\noalign{\smallskip}\tableline\noalign{\smallskip}
(a) CIBER/LRS intensity at 1.21$\, \mu$m & 792.09  & 465.29  & 406.42  & 417.74  & 388.17  \\
(b) Integrated starlight ({\it J}-band $>$13 mag) & 46.43  & 20.83  & 18.06  & 44.71  & 27.31  \\
(a) / (b) [\%] & 5.86  & 4.48  & 4.44  & 10.70  & 7.03  \\
\noalign{\smallskip}\tableline\noalign{\smallskip}
\multicolumn{6}{l}{{\bf Note.} --- Values in (a) and (b) are shown in nW/m$^2$/sr.}
\end{tabular}
\end{center}
\end{table*}

\subsection{Point Source Detections}
The flight data include spectra of stars which fell on the slits as bright horizontal lines as shown in Figure~\ref{NEP}. 
All detected horizontal lines were collected and masked when deriving the diffuse sky spectrum.
In this procedure, the images of point sources with $m_{\mathrm{Vega}}$({\it J}-band)$>$13 ($F_{0}(\mathrm{Vega}) = 1630 \,$Jy) were masked.
Remaining sources are also removed when we average the source-masked spectra with 3$\sigma $ clipping to derive the diffuse sky spectra.  
To estimate the contribution to the total sky brightness from Galactic stars fainter than the LRS detection limit, 
all {\it J}-band stars in the 2MASS point source catalog \citep{Skrutskie06} from $2^{\circ} \times 2^{\circ}$ regions centered at the each field are summed;
the 10$\sigma $ detection limit in 2MASS is 15.8 mag (Vega magnitude) in {\it J}-band.
To understand the completeness of the 2MASS-determined integrated starlight, 
the value determined from the 2MASS collection is compared with that computed using the SKY model which is designed to provide predictions for infrared Galactic star counts \citep{Cohen94}, 
and it is found that the calculated 2MASS integrated intensity was $\sim$90\% of the SKY model estimation.
The SKY model takes into account the flux from stars down to a much deeper limit of $\sim$30 mag which accounts for this difference.
We therefore corrected the calculated 2MASS integrated star light based on the SKY model estimation to account for stars fainter than 15.8 mag,
and the results are summarized in Table~\ref{LRSstar}.
The contribution to the sky brightness from residual Galactic stars was $\sim 5$\% besides NEP,
where the stellar contamination was largest ($\sim 10$\%) because of its low Galactic latitude.

\begin{figure}[htbp]
\begin{center}
\includegraphics[scale=0.35,angle=90]{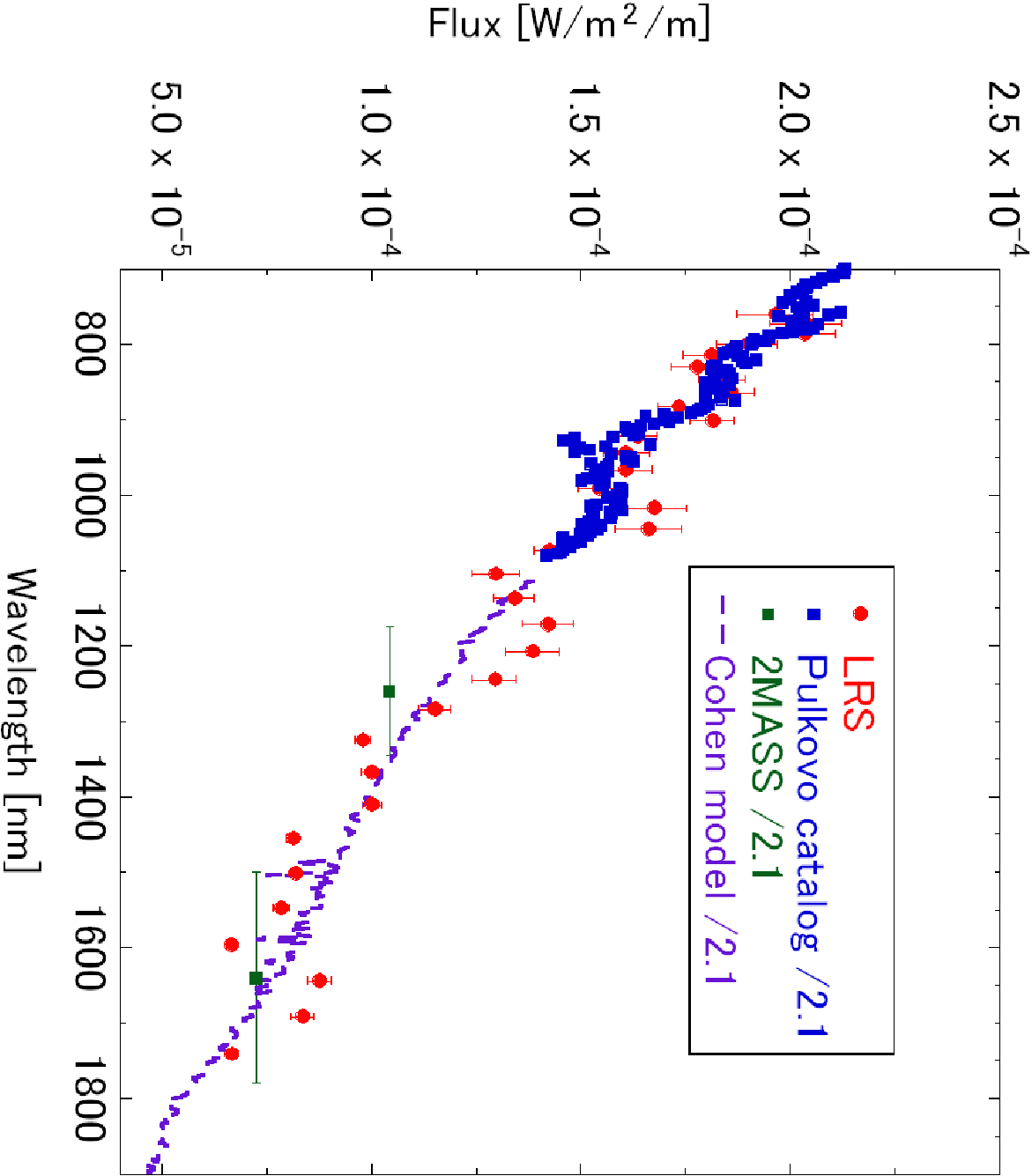}
\caption{Measured LRS spectrum of 42Dra; also plotted are the equivalent spectrum from the Pulkovo Spectrophotometric Catalog \citep{Alekseeva96}, 
a model spectrum \citep{Cohen03}, and photometric data from 2MASS \citep{Skrutskie06}.  
Because 42Dra was partially obscured by the slit mask, 
the curve has been scaled by a factor of 2.1 to normalize it about the absolutely calibrated spectra.
The LRS spectrum agrees with the other data, confirming the relative channel-to-channel calibration obtained in the laboratory.
The periodic spectral features evident in the Pulkova and LRS spectra are not reproduced by the model because of its less accuracy for the small spectral features.
\label{42dra}}
\end{center}
\end{figure}

The spectra of bright reference stars are used to check both the relative in-flight channel-to-channel response of LRS and the ground-based wavelength calibration.
Because the absolute pointing of the instrument is not known with sufficient precision, 
it is not possible to determine exactly where on the slits bright stars fell and so we are not able to use this information for absolute brightness calibration.
Figure~\ref{42dra} shows the measured LRS spectrum of 42 Draconis (42Dra, a mag$_H$=2.2, K1.5III star), the brightest star observed with the LRS during the flight.  
Also shown for comparison are the spectrum of 42Dra from the Pulkovo Spectrophotometric Catalog \citep{Alekseeva96}, the 2MASS fluxes \citep{Skrutskie06}, 
and a model of a K1.5III star $\alpha$-Boo \citep{Cohen03}. 
The LRS spectrum of 42Dra is 2.1 times fainter than the absolute brightness of the Pulkovo value; this difference is due to partial obscuration of the star by the slit mask. 
Although the model spectrum does not accurately reproduce spectral features observed with the LRS 
because the accuracy of small spectral features in the model spectrum is not very good,  
the LRS and Pulkovo spectra both exhibit such features in their regions of overlap.
Accounting for the accuracy of the model, the scaled LRS spectrum is consistent with the expected spectrum, 
confirming the relative response calibration obtained in the laboratory.

\section{RESULTS}
\subsection{Sky Brightness}
\begin{figure}[hbtp]
\begin{center}
\includegraphics[scale=0.35,angle=90]{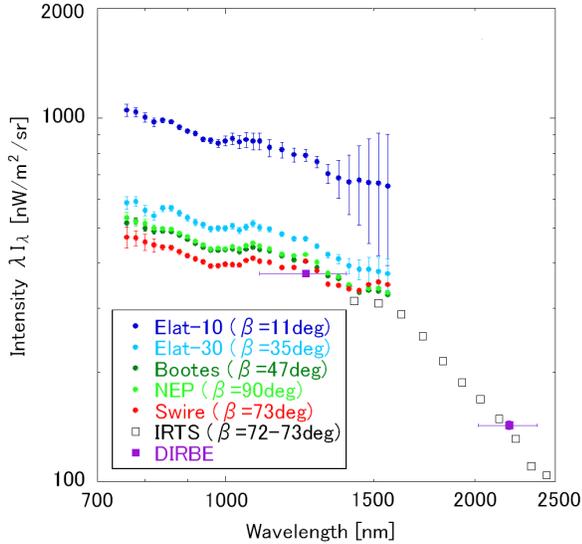}
\end{center}
\vskip -0.2cm
\caption{Measured spectra of the near-infrared sky brightness. Colored circles indicate the CIBER/LRS data used in this study, 
open squares indicate IRTS data averaged at 72$^{\circ }$-73$^{\circ }$ ecliptic latitude from \citet{Matsumoto96},
and purple squares indicate darkest DIRBE data from \citet{Hauser98}.
The error bars denote the combination of statistical and systematic error due to subtracting airglow emission.  
The absolute calibration error is estimated as 
$\pm 10$\% for LRS \citep{Tsumura10}, $\pm 5$\% for IRTS \citep{Noda96}, and $\pm 1.6$\% for DIRBE \citep{Hauser98}.\label{sky}}
\end{figure}

Figure~\ref{sky} shows the spectra of the sky brightness for fields observed during CIBER's first flight.  
The Elat-10 and Elat-30 plots show the spectra after subtracting airglow emission; airglow emission in the other fields is negligible.  
The LRS sky spectra show the total brightness from zodiacal light, Galactic stars, and the EBL, 
and are consistent with previous observations with {\it COBE}/DIRBE \citep{Hauser98} and IRTS \citep{Matsumoto96} in the region of overlap.
The spectral shapes of the sky brightness in Figure~\ref{sky} are very similar to each other, irrespective of ecliptic latitude. 
A previously unreported broad absorption feature centered at approximately 0.9$\, \mu$m is present in each of these spectra. 

\begin{figure}[hbtp]
\begin{center}
\includegraphics[scale=0.35,angle=90]{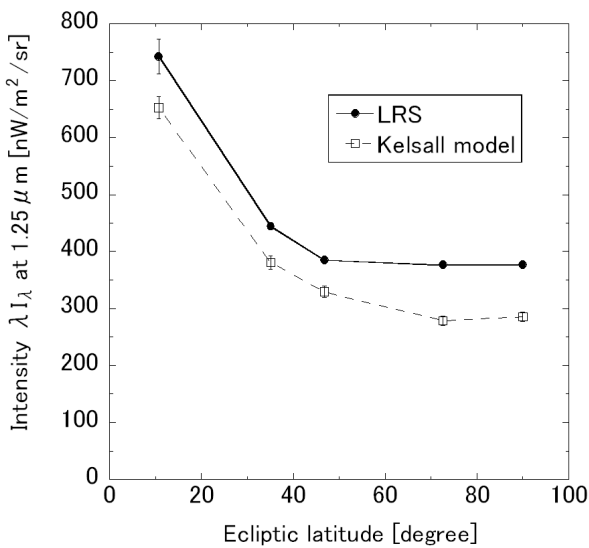}
\end{center}
\vskip -0.2cm
\caption{Ecliptic latitude dependence of the DIRBE all-sky zodiacal dust model \citep{Kelsall98} (open squares) 
and the LRS sky brightness (circles) after removal of the integrated Galactic starlight at $\sim 1.25 \, \mu$m.
The impact from the difference of the solar elongation or ecliptic longitude should be small, 
because the range of them in our observation fields is narrow ($<$30$^{\circ }$; see Table~\ref{flight}) to see the ecliptic latitude dependence of the zodiacal light.\label{eclp_lat}}
\end{figure}

Figure~\ref{eclp_lat} shows the ecliptic latitude dependence of the zodiacal light calculated using the DIRBE all-sky zodiacal dust model \citep{Kelsall98} 
and the observed sky brightness after removing the integrated Galactic star light at 1.25$\, \mu$m.
Except at high ecliptic latitude, the observed LRS brightness agrees with the ecliptic latitude dependence estimated by the model,
which shows again that our observation is consistent with the previous observations.
We note that the zodical light model departs from the data at higher ecliptic latitude.

\subsection{Zodiacal Light Spectrum}
\begin{figure}[hbtp]
\epsscale{0.92}
\plotone{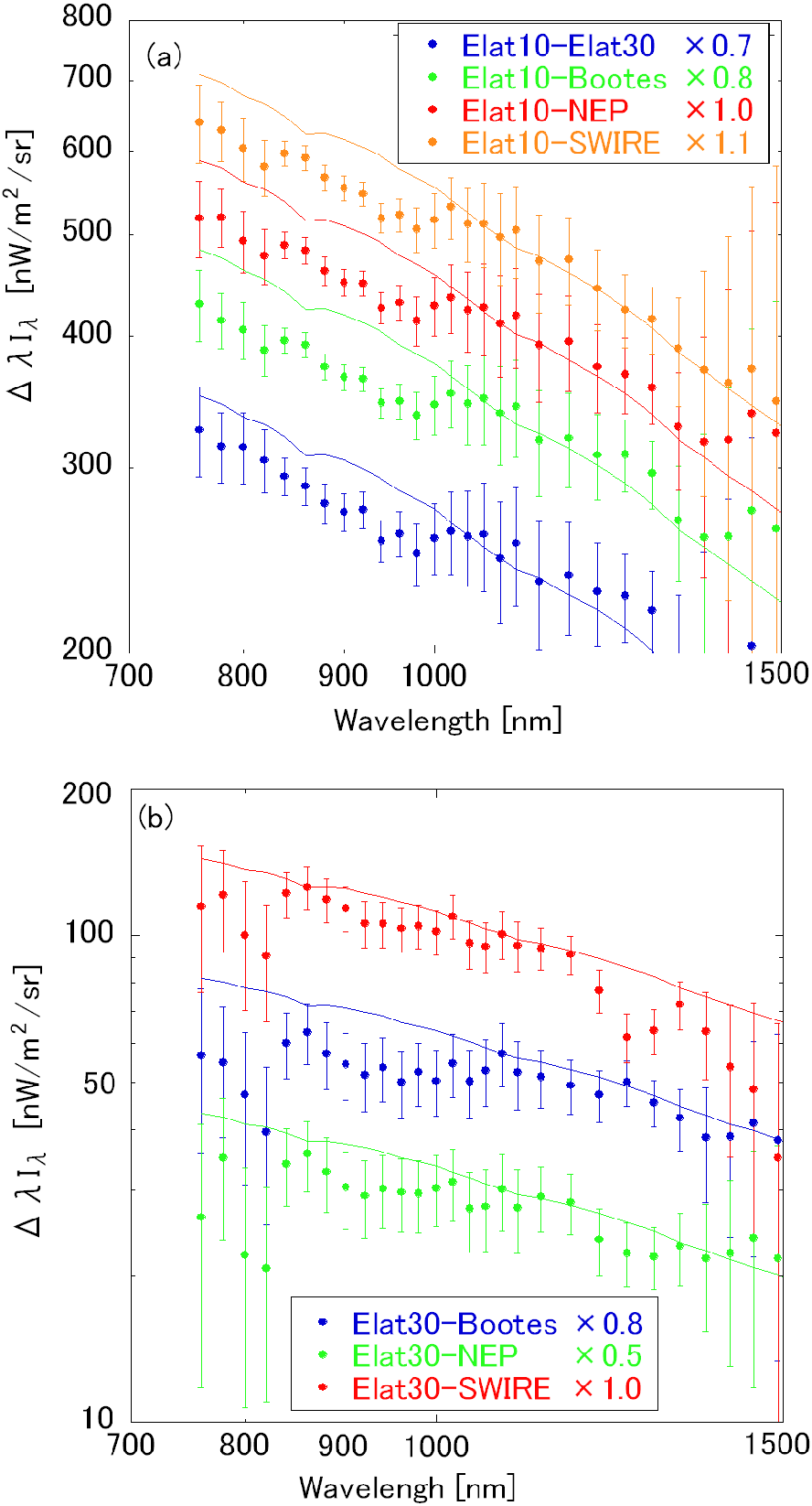}
\caption{Zodiacal spectra from various field difference combinations.
The figure (a) shows the difference between Elat-10 and the science fields, 
while the figure (b) shows the difference between Elat-30 and the science fields.
The curves are scaled to separate them for visualization.   
The solid lines denote the observed solar spectrum \citep{Gueymard02} normalized at 1.08$\, \mu$m. 
The difference spectra are noisier than the raw spectra as expected, but remove EBL and partially Galactic emission 
and show the absorption feature, confirming it is of zodiacal origin.\label{difference}}
\end{figure}

\begin{figure}[hbtp]
\vskip 0.5cm
\begin{center}
\includegraphics[scale=0.35,angle=90,clip]{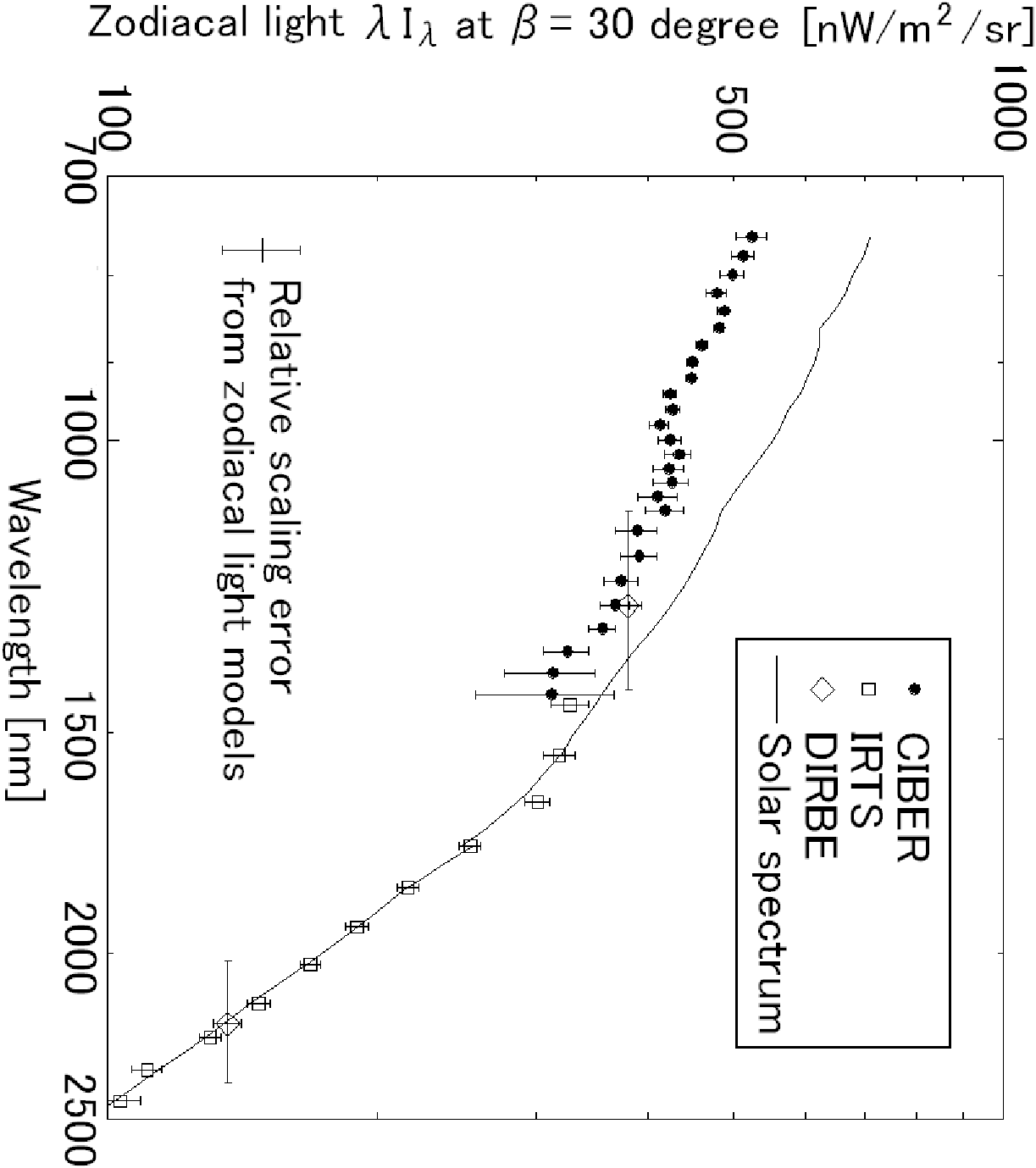}
\end{center}
\caption{Zodiacal light spectrum.
The filled circles show the CIBER/LRS data obtained by taking an average of the differences between Elat-10 and the other fields. 
The IRTS data \citep{Matsumoto96} are plotted in open squares and the curve denotes the observed solar spectrum \citep{Gueymard02} normalized to the IRTS data at 1.83$\, \mu$m.
The CIBER and IRTS data are normalized to the brightness expected at the Elat-30 field (see Table~\ref{flight} for specific coordinates) 
using the DIRBE zodiacal light model plotted in open diamonds \citep{Kelsall98}. 
Error bars on the LRS data denote the combination of statistical error and systematic error in subtracting airglow emission.  
Error bars on the DIRBE data show the CIBER-IRTS relative calibration error effecting the relative color calibration between CIBER and IRTS 
which is estimated to be 3.5\% from the color difference of the two different zodiacal light models \citep{Kelsall98, Wright98}.             
The relative scaling error due to the zodiacal light model is 10\% estimated from the difference of the absolute zodiacal intensity in the two models.\label{zodi}}
\vskip 0.5cm
\end{figure}

To separate residual astrophysical components from zodiacal light, the sky spectra observed in two fields are differenced; 
as shown in Figure~\ref{difference}, various combinations of such field differences show a similar spectral shape 
suggesting that the zodiacal light spectrum is largely isotropic. 
A joint CIBER-IRTS zodiacal light spectrum is produced by normalizing both data sets 
to the DIRBE all-sky zodiacal dust model \citep{Kelsall98} at $1.25\, \mu$m and $2.2 \,\mu$m using the bandpass response of DIRBE filters \citep{Hauser98b}.
Figure~\ref{zodi} shows the derived joint CIBER-IRTS zodiacal light spectrum with the solar spectrum \citep{Gueymard02}. 
The measured zodiacal light spectrum has a redder color than the solar spectrum,
and is generally similar to the unsubtracted spectra shown in Figure~\ref{sky}, though with much less signal to noise due to the differencing.  
Hereafter, LRS data at $>$1.4$\, \mu$m were removed because their error bars were too large because of the subtraction of the large airglow emission. 

\subsection{Reflectance of the Interplanetary Dust}
\begin{figure}
\epsscale{1.1}
\plotone{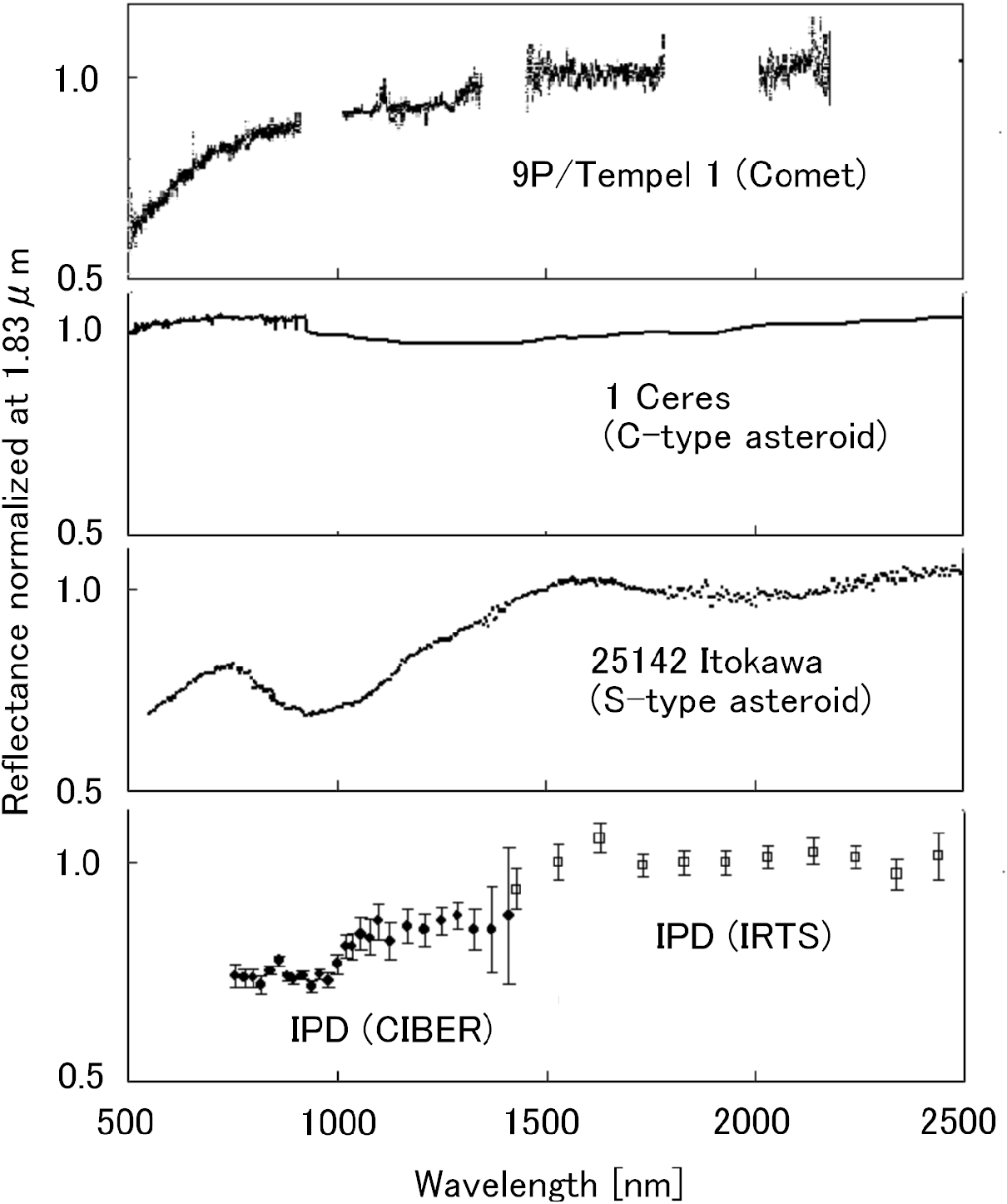}
\vskip 0.2cm
\caption{Reflectance of IPD normalized at 1.83$\, \mu$m. LRS and IRTS results are plotted together with example spectra of comets and asteroids from the literature.
The reflectance spectra of the LRS (filled circles) and IRTS (open squares) were obtained by dividing the zodiacal light spectra by the solar spectrum.
Curves show the reflectance of an S-type asteroid 25143-Itokawa from \citet{Binzel01}, a C-type asteroid 1-Ceres from \citet{Bus02},
and a comet 9P/Tempel-1 from \citet{Hodapp07}, respectively.\label{reflectance}}
\end{figure}

Figure~\ref{reflectance} shows the IPD reflectance spectrum, derived by dividing the LRS and IRTS spectra by the solar spectrum.
Note that the reflectance of zodiacal light as viewed from Earth is dominated by the IPD near Earth's orbit
because the radial density profile of IPD results in a contribution to observed intensity that falls off as $\sim r^{-2.5}$ \citep{Leinert98}.
The contribution of dust scattering at distances greater than 2.5 AU to the integrated zodiacal light intensity 
is estimated to be less than 3\% based on this radial density profile. 
The reflectance of an S-type asteroid 25143-Itokawa \citep{Binzel01}, a C-type asteroid 1-Ceres \citep{Bus02}, and a comet 9P/Tempel-1 \citep{Hodapp07} 
are also shown in Figure~\ref{reflectance} for comparison.
The C- and S-type asteroids are respectively the first and second most common types of asteroid in the asteroid belt.  
Unfortunately, due to the difficulty of such observations no spectra of old cometary dust trails at near-infrared wavelengths are available.
Instead we show the spectrum of ejecta material from comet 9P/Temple-1 after the impact of the deep impact probe \citep{Hodapp07};
this spectrum is of \textquotedblleft fresh\textquotedblright\ dust and other materials from the cometary nucleus and so is not an ideal comparison to the zodiacal reflectance spectrum.    

\section{DISCUSSION}
\subsection{Composition of the Interplanetary Dust}
We first detected the broad absorption feature in the IPD reflectance spectrum at 0.9$\, \mu$m in Figure~\ref{reflectance}.
This absorption feature is common in certain types of asteroids and meteorites, 
and it is thought to be caused by silicate compounds such as pyroxene and/or olivine,
which are the principal mineralogical constituent of IPD \citep{Bradley03}.
Variation of this feature over a wide range of ecliptic latitude (10-90$^{\circ }$) is not detected, 
which is consistent with mid-infrared observations \citep{Reach03}.

Ordinary chondrites have known absorption bands at $1 \, \mu$m and $2 \, \mu$m, evidence for the presence of pyroxene and olivine. 
The spectrum of 25143-Itokawa has been described as being similar to that of ordinary chondrite meteorites 
with additional reddening due to space weathering \citep{Binzel01}, a process that alters the state of surface materials exposed to a space environment.  
The exposed surface develops radiation damage, vapor- and sputter-deposited coatings, and melt products including agglutinates.
The reddening effect particular to space weathering is caused by a vapor coating containing nanophase-reduced iron (npFe$^0$) particles 
around each dust particle \citep{Pieters00, Nimura08}.
The spectrum of 25143-Itokawa is well matched to a model including 0.05\% npFe$^0$ reddening of ordinary chondrite \citep{Binzel01}, 
a value common to S-type asteroids \citep{Binzel01, Hapke00}. 
The similarity of the shape of the IPD reflectance measured by LRS to that of S-type asteroids is evidence that this process is relevant to the IPD responsible for producing the zodiacal light in the near-infrared.

\citet{Matsumoto96} reported some similarity between the spectrum of zodiacal light and S-type asteroids from the IRTS data.  
Our result is also consistent with the finding of a silicate feature in the mid-infrared zodiacal light spectrum \citep{Ootsubo98, Ootsubo09, Reach03},
though the peak position of the absorption band in the zodiacal light spectrum is slightly shifted from that of S-type asteroids.
This difference may be explained by differences in the chemical composition and/or some dependence on the size of the IPD particles.

\subsection{Distribution of the Interplanetary Dust}
\begin{figure}[hbtp]
\begin{center}
\includegraphics[scale=0.35,angle=90]{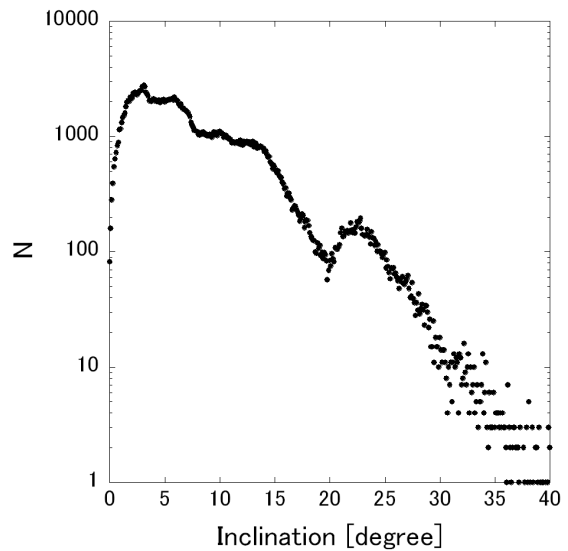}
\end{center}
\caption{Histogram of the asteroids in the asteroidal belt as a function of inclination of the orbit. 
The complete data set of the numbered asteroids from the JPL Web Site (http://ssd.jpl.nasa.gov/?sb\_elem) was used.\label{inclination}}
\end{figure}

As it is difficult to formulate a mechanism which propels asteroidal dust out of the ecliptic plane, 
recent dynamical analysis examining the ecliptic latitude dependence of the zodiacal emission at mid-infrared wavelengths 
suggests that the IPD arises mainly from a cometary population \citep{Nesvorny09}. 
Though the zodiacal emission is biased toward low-albedo dust from comets and C-type asteroids with albedo $\sim$0.05 \citep{Tedesco89},
the scattered light seen by the LRS may be biased by high-albedo S-type dust with albedo $\sim$0.2 \citep{Tedesco89}.
The result presented here can be consistent with the resent result favoring a cometary origin if high-albedo asteroidal dust is concentrated near the Earth.
Figure~\ref{inclination} shows the histogram of asteroids in the asteroidal belt as a function of their inclination, 
indicating that asteroids are distributed over 10$^{\circ }$-15$^{\circ }$ inclination at half maximum. 
Assuming that the asteroidal dust generated in the asteroidal belt falls into the sun purely by the Poynting-Robertson effect,
the estimated scale height of the dust cloud from the asteroid belt at 1 AU is $\sim$0.2 AU. 
This thickness is comparable with the estimated IPD cloud scale height of the DIRBE all-sky zodiacal dust model \citep{Kelsall98}. 
Cometary dust may be distributed at distances greater than 0.2 AU from the Earth, 
if the ecliptic latitude dependence of the zodiacal light is masked by the nearest high-albedo asteroidal dust.

The spatial distribution of asteroid types within the asteroid belt is non-uniform.
Although S-type asteroids are the second largest population within the asteroidal belt, they predominantly occupy the inner portion of the belt.
The relative fraction of S-type asteroids at heliocentric distances out to 2.5 AU is no less than 60\% \citep{Mothe03}.
Using this fractional abundance and assuming that dust production rate of each asteroid type is proportional to the square of the number density, 
the dust ratio of S-type to C-type asteroids is expected to be 5:3.
Furthermore, the C-type asteroids have a smaller albedo ($\sim$0.05) compared to S-type asteroids ($\sim$0.2) \citep{Tedesco89}.
The combination of asteroid type ratio and the albedo ratio suggests that 
S-type dust dominates over C-type dust in the observed zodiacal light spectrum by a fraction of 6 and 7.

These estimates are qualitatively consistent with the interpretation that the CIBER/LRS zodiacal light spectrum shows the existence of the S-type asteroidal dust particles,
and that they are largely responsible for the scattered zodiacal light spectrum observed from Earth.
Our result, however, does not rule out an important contribution from comets. 
We are unable to state the exact dust fraction from comets versus asteroids as the high albedo dust near the Earth's orbit dominates the observed spectrum.
Furthermore there is still the possibility that cometary dust might also show the pyroxene/olivine feature at $0.9 \, \mu$m.
Pyroxene and olivine have been confirmed in the mid-infrared spectrum of the cometary dust \citep{Lisse06} 
and in the cometary dust samples returned by the Stardust spacecraft \citep{Zolensky06}, 
but there is no evidence of these absorption features in the near-infrared spectrum. 
Spectroscopic observations of the cometary dust trails near 1$\, \mu$m will be necessary to constrain the cometary contribution to the observed LRS spectrum,
and to estimate the fraction of dust from comets in the IPD.
A measurement of near-infrared and mid-infrared spectral features obtained in a deep space mission traveling to the outer solar system could map the composition of IPD throughout the solar system,
and trace its origins to asteroidal and cometary parent populations \citep{Matsuura02, Cooray09}.

\acknowledgments
\section*{ACKNOWLEDGMENTS}
This work was supported by KAKENHI (20$\cdot$34, 18204018, 19540250, 21111004, and 21340047)
from Japan Society for the Promotion of Science (JSPS) and the Ministry of Education, Culture, Sports, Science and Technology (MEXT),
and NASA APRA research grant (NNX07AI54G, NNG05WC18G, NNX07AG43G, and NNX07AJ24G). 
We acknowledge the dedicated efforts of the sounding rocket staff at NASA Wallops Flight Facility and White Sands Missile Range. 
We also acknowledge the engineers at the Genesia Corporation for the technical support of the CIBER optics. 
We thank Dr. Allan Smith, Dr. Keith Lykke, and Dr. Steven Brown (the National Institute of Standards and Technology) for the laboratory calibration of LRS.
We also thank Dr. Hasegawa Sunao (ISAS/JAXA), Dr. Ishiguro Masateru (Seoul National University), Dr. Ootsubo Takafumi (Tohoku University), 
Dr. Noguchi Takaaki (Ibaraki University), Dr. Pyo Jeonghyun (KASI), and Dr. Carey Lisse (Johns Hopkins University) for discussions and comments about IPD,
and Dr. Martin Cohen (UC Berkeley) and Dr. Yamamura Issei (ISAS/JAXA) for comments about stellar spectrum.
K.T. acknowledges support from the JSPS Research Fellowship for the Young Scientists, M.Z. acknowledges support from a NASA Postdoctoral Fellowship, 
and A.C. acknowledges support from an NSF CARRER award. \hspace{1000bp}

\end{document}